\documentclass[article,12pt,onecolumn]{IEEEtran}
\pdfoutput=1

\usepackage[superscript]{cite}
\usepackage{hyperref}
\usepackage[pdftex]{graphicx} 
\usepackage[cmex10]{amsmath}
\usepackage[tight,footnotesize]{subfigure} 
\usepackage[format=hang]{caption} 
\usepackage{xcolor} 

\title{Characterizing Persistence and Disparity of Covid-19 Infection Rates with City Level Demographic and Regional Features}
\author{Emi Aoki, Arkajyoti Sinha, Charles Thompson, and Kavitha Chandra}

\providecommand{\keywords}[1]
{
  \textbf{Key words:} #1
}

\begin{document}

{\let\newpage\relax\maketitle}

\section*{\noindent\raggedright\textbf{ABSTRACT}}
\normalsize

\noindent\raggedright\subsubsection*{\textbf{BACKGROUND AND OBJECTIVE}}
The design of data-driven dashboards that inform municipalities on ongoing changes in infections within their community is addressed in this research. An approach for  distilling data provided at the state level into meaningful information for individual municipalities to identify and recommend suitable interventions for the affected population is presented. Daily reports of Covid-19 infections published by the state of Wisconsin as the initial surge in the pandemic ensued during the October 2020 to September 2021 time frame is considered as a case study. Of particular interest is the identification of regions and population groups distinguished by race and ethnicity that may be experiencing a disproportional rate of infections over time. This study integrates the municipality-level daily positive cases that are disaggregated by race and ethnicity and population size data derived from the US Census Bureau. The goal is to present timely data-driven information in a manner that is accessible to the general population, is relatable to the constituents and promotes community engagement in managing and mitigating the infections.

\raggedright\subsubsection*{\textbf{METHODS}}
A statistical metric referred to as the rank difference and its persistence over time is used to capture the disproportional incidence of Covid-19 positive cases on particular race and ethnic groups in relation to their population size. Each of the 190 Wisconsin municipalities and its four population groups is characterized by the rank-difference time series and a skewness measure. A persistence index is derived to identify regions that continually exhibit positive rank differences on a daily time scale and indicate disparity in disease incidence. This index, alongside a comparison of the relative difference in infections experienced by Black/African American and Hispanic/Latino groups compared to the majority White population in each of the regions is mapped to a dashboard for each municipality.

\raggedright\subsubsection*{\textbf{RESULTS}}
The analysis leads to the identification that several municipalities in Wisconsin that are located in regions of low population and away from the denser urban centers are those that continue to exhibit disparity in the infection rates for Black/African American and Hispanic/Latino population groups. Examples of a dashboard that can be utilized to capture both aggregate level and temporal patterns of Covid-19 infections are presented.

\raggedright\subsubsection*{\textbf{CONCLUSIONS}}
Statistical measures that describe incidence of Covid-19 infections in relation to the population size of a locality are derived and proposed as insightful information for presentation on dashboards to promote community engagement and timely interventions. These measures are particularly sensitive in revealing the disparate effect of the disease on minority groups in regions with low to moderate population sizes.

\keywords{Covid-19, Wisconsin, rank-difference, race and ethnicity, dashboards}

\IEEEpeerreviewmaketitle

\newpage
\section*{\raggedright\textbf{INTRODUCTION}}
The onset of the Covid-19 pandemic resulted in the creation of many online platforms for visualizing data on the number of people who were infected, hospitalized, or succumbed to the disease across the nation and the world.  Even as the areas where a rapid rise in infections were broadcast on television and social media, recommendations of nonpharmaceutical interventions (NPIs) such as masking and isolation for mitigating the spread of the disease were met with skepticism in certain regions and among some groups of people. These events have made evident the need for the public to have a more nuanced understanding of the state of the disease in a manner that is relatable to their own personal circumstances and in  context of their community needs.  The metrics of infectious disease experts such as the reproduction number and related  messaging for public cooperation to drive this number down may not be as insightful amid mass confusion and lack of basic amenities in the early days of the pandemic.  Moreover, the implementation and managing of NPIs \cite{pageGostin2020} is challenging based not only on the country but also on occupational, economic, and regional demographics such as for frontline workers, unemployed and senior citizens and those residing in rural areas. Gilmore \emph{et al.}
 \cite{paperGilmoreEtAl2020} conduct a review of literature on past epidemics and note the importance of social and community engagement and responses in reaching marginalized populations and supporting equity-informed responses. The effectiveness of public health messages that were clear, credible and presented in a language understood by target groups has been identified by Lawes-Wickwar \emph{et al.} \cite{paperLawesWickwareEtAl2021} in prescribing vaccine interventions. 

The disproportionate impact of Covid-19 infections on people belonging to particular race and ethnic groups has become evident from various data-driven studies since the onset of the pandemic. The Centers for Disease Control and Prevention (CDC) has compiled some of this evidence, summarizing by race and ethnicity, Covid-19 cases, hospitalization, and death rate ratios \cite{pageCDC2022covidCaseHospDeathByRaceEth}. Based on relative population sizes, Black or African American (BAA), American Indian or Alaska Native (AIAN), and Hispanic or Latino (HL) populations were found to have experienced nearly two or three times the detrimental effects of Covid-19 compared to the White (W) population. 
Several research studies that pertain to connections between demographic and/or geographic
features and Covid-19 infections have been published. These studies typically characterize and compare the impact of Covid-19 in terms of rates per 100,000 of the considered population. The Community Resilience and Response Task Force (CRRTF) report from the state of Wisconsin \cite{pdfRacialEquityInWI2021} provided information on the state’s racial disparity in Covid-19 cases, hospitalizations, and deaths using state-wide data. They recognized a need for addressing racial structures of the community to effectively mitigate the pandemic. Covid-19 outcome rates per 100,000 people has shown the disproportional effect on communities of color. However, reports at this level of aggregation often fail to resonate in regions of moderate to low population, many of which exist in Wisconsin particularly in the northern and western areas of the state. 
Rast \emph{et al.} \cite{paperRastEtAl2020Milwaukee} and Egede \emph{et al.} \cite{paperEgedeEtAl2020WICpRace} identify the association between the distributions of BAA and/or HL population and Covid-19 cases in Southeast Wisconsin in the early stage of the outbreak. Rast \emph{et al.} \cite{paperRastEtAl2020Milwaukee} finds that areas with high BAA population percentage form clusters of large number of confirmed cases compared to White residents. Egede \emph{et al.} \cite{paperEgedeEtAl2020WICpRace} hypothesize that racial and ethnic disparities in Covid-19 cases are likely to be due to structural racism. In the regions analyzed in these papers, BAA and/or HL population percentage was in the range of twenty to thirty percent which is relatively high compared to most other regions of Wisconsin. Overall, in the state of Wisconsin, the percentage of BAA and HL population percentages are 7.3\% and 7.0\% respectively as of 2018.  In this paper, measures for assessing the number of Covid-19 positive test cases in relation to the population ranking of the 190 municipalities of Wisconsin are presented. The focus of this research is to highlight measures of community impact that are understandable by the general population. Such measures may be useful in improving the adoption of interventions proposed at the city, state or national level.

\section*{\raggedright\textbf{METHODS}} 
\label{sec:methods}

\subsection*{\textbf{Data Collection: Covid-19, Demographic and Geographical Data Sources }}
The analysis presented in this work is based on information drawn from open access data sets published by the State of Wisconsin Health Services (WI-DHS)\cite{dataWICOVID19HistCntsumV2} and the United States Census Bureau from
which the American Community Survey (ACS)\cite{dataCensusB03002RACE} and cartographic boundary files\cite{dataCensusCartographicBoundaryCounty}$^{,}$\cite{dataCensusCartographicBoundaryCousubdiv} are retrieved. 
The WI-DHS8 data provides a daily report of the number of confirmed or likely Covid-19 positive cases in each of 190 municipalities in the state of Wisconsin. The report also disaggregates the case count by race and ethnicity considering the following seven groups: White (W), Black/African American (BAA), Asian or Pacific Islander (A), American Indian or Alaska Native (AIAN), multiple races (MO), Unknown race (UNK), and Hispanic/Latino (HL).
The population sizes of these groups in each of these municipalities are obtained from the ACS
2015-2019 dataset that also provides population counts by race and ethnicity \cite{dataCensusB03002RACE}. 
In this paper, the four racial and ethnic groups W, BAA, HL, and OTH are studied. The OTH group combines the Asian, HPI, and AIAN population.
The cartographic boundary files retrieved from the U.S. Census Bureau contain county \cite{dataCensusCartographicBoundaryCounty} and municipal-level \cite{dataCensusCartographicBoundaryCousubdiv} geospatial data represented by a list of longitude and latitude variables that allow visualization of the spatial patterns of Covid 19 effects on the map of Wisconsin.
By combining municipal-level Covid-19 positive cases and the data on population sizes of these regions, the impact of the outbreak on the four groups considered is examined. The geographical regions affected are visualized by mapping the findings onto the Wisconsin county map.

\subsection*{\textbf{Motivation}}

The pattern of infection rates of the BAA and HL subjects in comparison with the majority white population over the duration of one year from October 1, 2020 to September 30, 2021 is analyzed. The goal is to identify geographic regions and population groups that may have experienced a disproportional rate of infections relative to their population size. Also of interest is the derivation of data-derived indicators that can isolate regions and groups experiencing persistent disproportionate impact of Covid-19 over time.

This is motivated by the initial observation of how Covid-19 positive cases aggregated over the entire state of Wisconsin evolved over the year for each of these groups. 

A 7-day moving average of the daily counts of positive cases is taken as advised by the WI-DHS \cite{pageWIDHSdataInfo} and shown in Fig. 1. The first peak which occurred around November-2020 (day 54) shows that 0.16 \% of HL population in Wisconsin were affected compared to 0.12 \% for group W and 0.10 \% for BAA and OTH groups. The second peak around April-2021 (day 202) shows that the BAA population experienced higher rate of infection as shown by the orange-colored graph and this trend persisted for the BAA group for the rest of the year.

\begin{figure}
    \centering
    \includegraphics[scale=0.7]{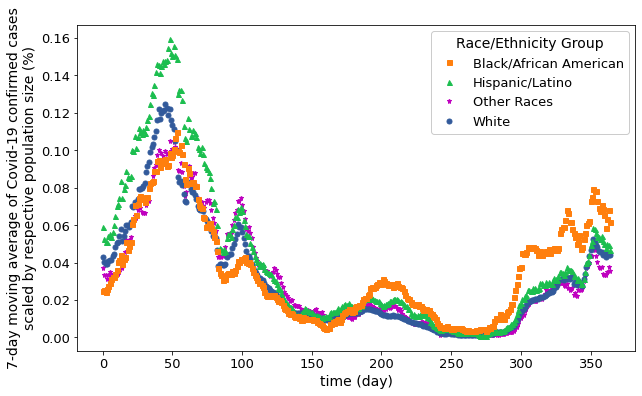}
    \caption{7-day moving average of new confirmed or probable Covid-19 cases for the four race and ethnicity groups. The vertical axis represents the counts scaled by respective group population size (\%). The Other races category combines Asian, Pacific Islander, and American Indian and Alaskan Native population. Day 1 corresponds to Oct. 1, 2020, and day 365 corresponds to Sep. 30, 2021.}
    \label{fig:7dCp_raceEth_ttlacross190cities_scaledByPoplSize}
\end{figure}

The rest of this paper is organized as follows. In the next section, the application of a rank difference metric for identifying municipalities and population groups that are adversely affected is discussed. The statistics of the rank difference time series that provide intuitive evidence of disproportionate cases are presented in the Rank Difference Statistics section. Finally, in the section entitled A Municipality Dashboard Application two examples of how the statistics may be presented to the occupants of the region is discussed.

\subsection*{\textbf{Ranking Population Size and Covid Cases}}
The data representing Covid positive cases for $M=190$ municipalities and $N=365$ days and $K = 4 $ population groups is organized in a three-dimensional matrix $S(i,j,k)$ for $i=1,2..,M$, $j=1,2...N$ and $k=1,2,3,4$ where $k=1$ is BAA, $k=2$ is HL, $k=3$ is OTH and $k=4$ is group W.

Assuming that the population size of the $M$ regions does not change significantly over the year for each of the four groups, the municipalities are ordered in decreasing group population size and assigned successively increasing integer rank values. For example, the cities of Milwaukee and Madison being the largest regions assume rank values of 1 and 2 respectively. This function will be denoted as $R_p[i_{ok}]$, $i_{ok}=1,2...M$ $k=1,2,3,4$ and it has a range from 1 to 190. The index $i_{ok}$ refers to an ordered sequence of the 190 municipalities based on the population size of the $k^{th}$ group.

A similar rank ordering is carried out by ordering the Covid-19 data in S(i,j,k) in descending order for each instant in time $j$ and for each of the $k$ groups. This yields a function $R_c[i_{ok},j]$ that tabulates for each municipality index $i_{ok}$ of the $k^{th}$ group its corresponding rank in Covid cases as a function of time. 

Figures 2(a) and (b) show a graph of $R_p(i_{o1})$ overlaid with $R_c(i_{o1},j=54)$ for BAA case and $R_c(i_{o4},j=54)$ for group W as a function of their respective ordered municipality indices $i_{o1}$ and $i_{o4}$. The blue line remains invariant with time and represents the increasing rank value as a function of decreasing $k^{th}$ group population size. The orange symbols capture the Covid rank on the $54^{th}$ day of the dataset. It is evident that for group BAA a significant number of cities move up in rank (i.e. assume lower values), as observed by their dispersion below the blue line. This is indicative that the Covid positive case count this group is experiencing is significantly higher than expected based on their population size. In contrast, Fig. 2(b) shows these metrics for group W, where the dispersion from the blue line is considerably smaller in size. 

These observations lead to the definition of a rank difference metric to capture the state of infections as a function of time for each of the four population groups. 
\begin{figure}[h]
    \centering
    \subfigure[]{\includegraphics[scale=0.6]{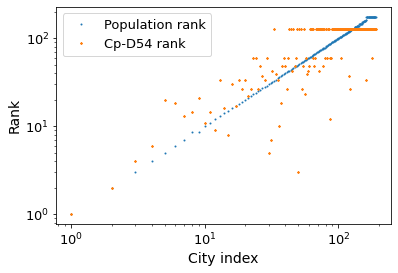}}%
    \subfigure[]{\includegraphics[scale=0.6]{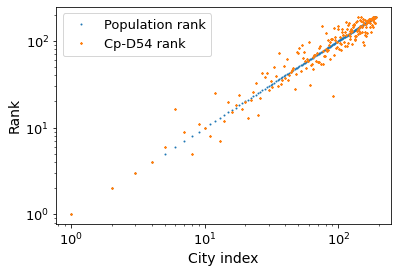}}
    \caption{Population size rank, $R_p[i_{ok},]$ (a) for $k=1 (BAA)$ and (b) for $k=4$ (W) compared with Covid-19 case rank $R_c[i_{ok}, j=54]$.}
    \label{fig:cityIndex_ranks}
\end{figure}
\newline
\subsubsection*{\textbf{Rank Difference Statistics}}
The rank difference (\(rd\)) is calculated as the difference between $R_p(i_{ok})$ and $R_c(i_{ok}, j)$ for each of the 190 municipalities and for each instant of time $j=1, \dots ,365$ and the four population groups indexed by $k$. 
\begin{equation}
    rd\,[j,i_{ok}] = ~ R_p(i_{ok}) - R_c(i_{ok},j) 
\end{equation}

The time series $rd(j,i_{ok})$ captures the state of Covid 19 infections in each of the 190 municipalities in relation to its $k^{th}$ population size. It ranges from -190 to +190. A positive value of $rd$ and in particular its persistence in the positive regime is indicative of regions where disproportionate levels of infections have continued to occur over time. 

The rank difference statistic is proposed as a useful measure of the regional state of infections as it can be visualized over time and also summarized in an intuitive way that may appeal to the population of the region.

The next section presents statistical measures of the rank difference time series for each municipality that can be recorded over successive time intervals of observations and applied to inform the public on the state of the disease in their locality.
\newline

\subsubsection*{\textbf{Persistence and Skewness Measures of Rank Difference Metric}}

Each of the 190 municipalities exhibit random variations in the rank-difference $rd$ as a function of time. The greater the positive value of $rd$ and its continued behavior in this regime, the more significant is the effect of Covid 19 infections for that region.

The degree of persistence is captured by defining a regime $ T_{rd} : T_{min} < rd < T_{max} $ and counting the number of instances of $rd$ that reside in this regime over a specified time interval. An indicator variable $I\,[j,i_{ok}] = 1 $ if $rd\,[j,i_{ok}] \in T_{rd} $ and zero otherwise. The persistence index for location identified by $i_{ok} $ is defined as:
\begin{equation} 
per\,[i_{ok}] = \frac { \sum_{j=1}^{365} I\,[j,i_{ok}] } { 365 } \times 100
\end{equation} 
considering a year of observations. It ranges from $0$ to $100 \%$.

\begin{figure}[h]
    \centering
    \includegraphics[scale=0.35]{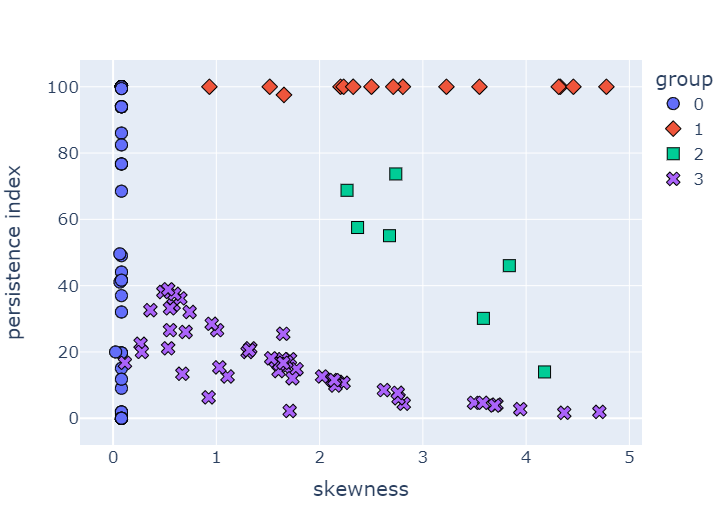}
    \caption{Skewness and persistence index for the BAA population in 190 municipalities.}
    \label{fig:stats_pers_skew_kurt_BAA}
\end{figure}

The shape of the distribution of $rd$ for each municipality provides an aggregate representation of the temporal observations. The third central moment of $rd$ noted as the skewness is of interest to capture the asymmetry and in particular the relative skew towards the positive values of $rd$. The adjusted Fisher-Pearson coefficient of skewness [ref] is applied in this analysis.  
\newline
Fig.~\ref{fig:stats_pers_skew_kurt_BAA} captures for the BAA group ($k=1$), the 156 municipalities that were found to have positive valued skewness in their rank difference time series as a function of corresponding persistence index of $rd$ time series. The threshold regime $ T_{rd} : 0 < rd \le 190 $.

This characterization separates the municipalities into four groups noted by index 0,1,2 and 3 and represented by different colored symbols. Group 0 includes eighty cities that have almost symmetric distributions of $rd$ and a persistence range from 0 to 100 percent. Group 1 identifies fifteen cities with $per[j,i_{o1}] \geq 90\%$ and a wide range of positive valued skewness from 1 to 5. Group 2 include seven cities with moderate and lower levels of persistence and skewness ranging from 2-4. Group 3 are fifty-four cities demonstrating a functional dependence of inverse proportionality between the persistence index and skewness metric.  

\begin{figure}
    \centering
    \includegraphics[scale=0.6]{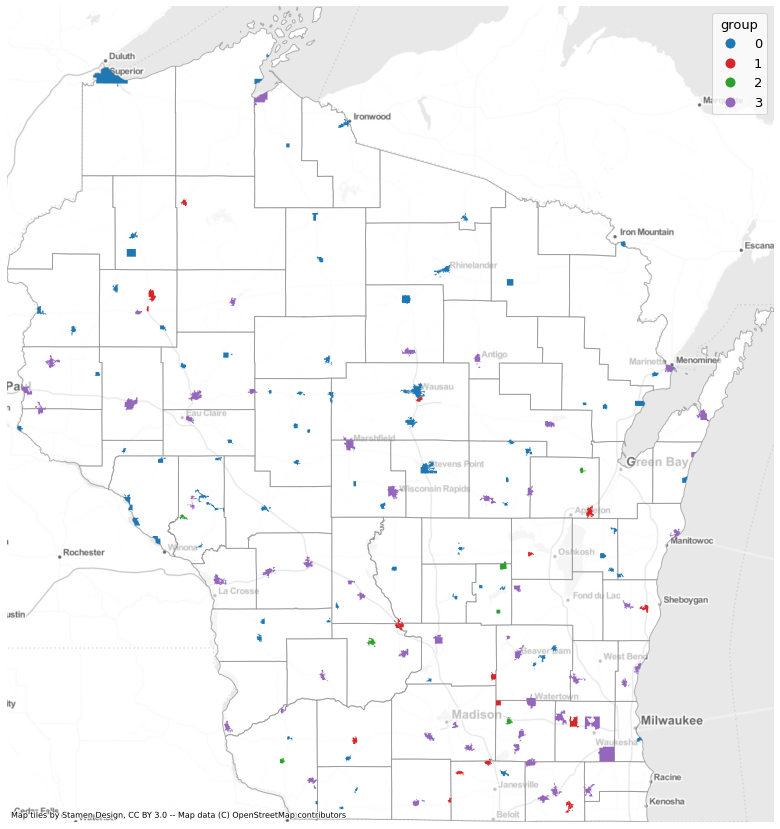}
    \caption{Mapping four groups shown in Fig.~\ref{fig:stats_pers_skew_kurt_BAA}  onto Wisconsin county map.}
    \label{fig:map_persistence_BAA_0-50}
\end{figure}

Fig.~\ref{fig:map_persistence_BAA_0-50} shows the Wisconsin state county map on which are marked the four groups of municipalities distinguished by the persistence index and skewness metric. Regions from group 0 marked in blue are primarily located in northern and western part of the state, which is much less populated than the southeastern region where the most populous cities are located. The BAA region population in these regions is also correspondingly lower being less than 100 people per municipality \cite{pdfRacialEquityInWI2021}. Most of the cities in group 1 and 2 are located in southern, and southeastern regions, where both BAA and HL populations are moderately higher than that of group 0. Group 3 municipalities shown in purple color are spread out across the state and have a large variation in their population sizes. 

\subsection*{\textbf{A Municipality Dashboard Application}}
A dashboard application of the statistical measures derived in the previous sections for presentation to the public is discussed in this section. The goal is to show the impact of Covid-19 infections on minority populations relative to the majority population in the region. The relative percentage difference in Covid-19 positive cases between BAA, HL and OTH groups with reference to the majority group W group is calculated as follows. Let $Cp_{ik}$ denote the number of positive infections recorded for group $k$ in municipality $i$ and $P_{ik}$ the population size of group $k$ in region $i$. The ratio $\frac{Cp_{ik}}{P_{ik}}$  is compared to $\frac{Cp_{i4}}{P_{i4}}$ to determine the relative change of group $k$ with respect to group W for $k=4$. 
The percent relative change is computed as follows~\cite{paperTornqvist1985relativeChange}
\begin{equation}
    H(i,k) = 100 \, \left [ \frac{y_{ik}-x_{i4}}{x_{i4}} \right ] 
\end{equation}
where $y_{ik}=Cp_{ik}/P_{ik}$ for $k=1,2,3$, $x_{i4}=Cp_{i4}/P_{i4}$. Values of $H(i,k)$ around zero indicate that Covid-19 infections are proportional to population sizes whereas positive values of $H(i,k)$ identify municipalities where there is a disproportionate infection rate of the group referenced by index $k$. 

\begin{figure}[h]
    \centering
    \includegraphics[scale=0.55]{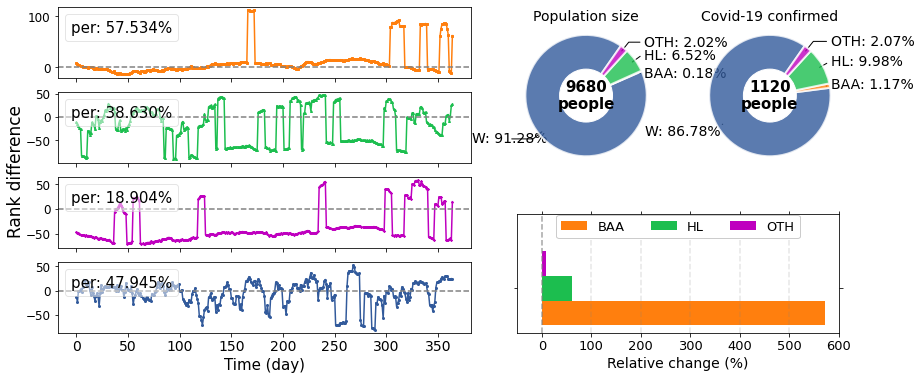}
    \caption{Dashboard for one of cities in group 2}
    \label{fig:dashboard_Reedsburg}
\end{figure}

Fig.~\ref{fig:dashboard_Reedsburg} shows an  example of a dashboard for one of cities classified as group 2 with respect to persistence index and skewness in Fig. \ref{fig:stats_pers_skew_kurt_BAA}. This dashboard summarizes population and Covid-19 cases by race and ethnicity (pie charts), $rd\,[j,i_{ok}]$ time-series for $k=1,2,3$ corresponding to BAA, HL and OTH population groups, their persistence index and the relative change measure given by $H(i,k), ~~k=1,2,3$. This case highlights a BAA population of 0.18\%, that shows 1.17\% of this population experiencing disproportional number of Covid-19 infections. The relative change from group W is nearly 550\% that further emphasizes this finding. 

Three special cases are observed for relative change values. The first special case is a city with zero population size and zero Covid cases ($Cp_{ik}=0$ and $P_{ik}=0$); thus, relative change is undefined and marked with a cross. The second case is a city with non-zero Covid cases while its population size is zero according to the ACS data ($Cp_{ik}\neq 0$ and $P_{ik}=0$), which is marked with a star. The last case is a city when the number of Covid-19 cases is larger than non-zero population size ($Cp_{ik} >  P_{ik}$ and $P_{ik}\neq0$), which is marked as a triangle.

The dashboard shown in Fig.~\ref{fig:dashboard_Reedsburg} indicates that all three racial/ethnic  groups $k=(1,2,3)$ are disproportionately affected compared to the W group as shown although their total population size is less than 10\% of the region's population. This finding might not be reported at the state level in general as their city's population size is less than 100,000.

\begin{figure}[h]
    \centering
    \includegraphics[scale=0.55]{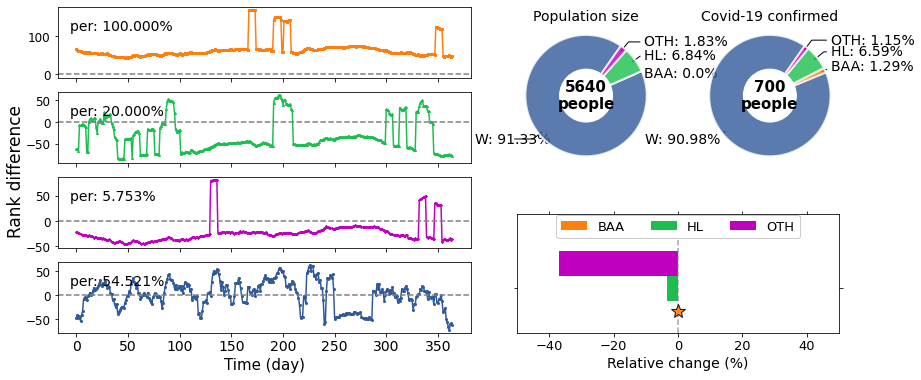}
    \caption{Dashboard for one of cities in group 1, which exhibits one of special cases for relative change.}
    \label{fig:dashboard_Edgerton}
\end{figure}

Fig.~\ref{fig:dashboard_Edgerton} shows a case when a municipality experiences one of special cases of relative change for BAA data. The star mark indicates that BAA population size is zero although the number of Covid-19 case is non-zero. This probably increases rank difference values, resulting in 100\% persistence index for BAA while the Covid cases for HL and OTH groups are lower compared to W group.

All cities in group 1 are found to show special cases (star or triangle) for BAA relative change.

As seen in Fig.~\ref{fig:dashboard_Reedsburg} and~\ref{fig:dashboard_Edgerton}, each region is likely to have unique trends due to various factors contributing to the transmission of the infectious disease. Thus, the community-level data visualization as demonstrated here can be helpful to inform people on their own community's situation and help mitigate the pandemic.

\section*{\raggedright\textbf{CONCLUSIONS}}
\label{sec:conclusion}
This research presented a statistical metric referred to as a rank difference to capture the disproportional number of Covid-19 cases experienced by Black/African and Hispanic/Latino population groups. 
An analysis of city-level Covid-19 positive cases experienced by four different race/ethnic groups was presented for 190 municipalities in the state of Wisconsin. The data analyzed was integrated from the Covid-19 positive test cases provided by the Wisconsin State Health Services and the population sizes given by the US Census Bureau. The rank difference time-series that showed persistence in a positive range of values for over a year of analysis were found to be in a subset of municipalities with a relatively low population of BAA and HL groups. The persistence of the rank-difference over time was captured by a persistence index statistic. The distribution of this variable allows the identification of the cities that persist in the positive range for large periods of time. A dashboard summarizes the data analysis and its findings for each of municipalities so that people in each community can be informed of their situation to help build effective mitigation strategies. Future work will address the application of the statistical metrics identified in this work to other states in the US.

\section*{\raggedright\textbf{Acknowledgement}}
The authors acknowledge partial support of this research from the National Science Foundation under Grant No. \#2128749. Any opinions, findings, and conclusions or recommendations expressed in this material are those of the authors and do not necessarily reflect the views of the National Science Foundation.

\ifCLASSOPTIONcaptionsoff
  \newpage
\fi

\newpage
\renewcommand{\refname}{\raggedright\textbf{References}}


\begin{thebibliography}{10}
\bibitem{pageGostin2020}
Gostin LO. {Could - Or Should - The Government Impose A Mass Quarantine On An
  American City ?} {\it Health Affairs Blog.} 2020.

\bibitem{paperGilmoreEtAl2020}
Gilmore B, Ndejjo R, Tchetchia A, et al. {Community engagement for COVID-19 prevention and control: a rapid evidence synthesis} {\it BMJ Global Health.} 2020;5:e003188.

\bibitem{paperLawesWickwareEtAl2021}
Lawes-Wickwar S, Ghio D, Tang MY, et al. {A Rapid Systematic Review of Public Responses to Health Messages Encouraging Vaccination against Infectious Diseases in a Pandemic or Epidemic} {\it Vaccines.} 2021;9:72.

\bibitem{pageCDC2022covidCaseHospDeathByRaceEth}
{Centers for Disease Control and Prevention}. {COVID-19 Hospitalization and Death by Race/Ethnicity.} 2020.
\newblock Available at: \url{https://www.cdc.gov/coronavirus/2019-ncov/covid-data/investigations-discovery/hospitalization-death-by-race-ethnicity.html}. Accessed on February 25, 2022.

\bibitem{pdfRacialEquityInWI2021}
{WI Community Resilience and Response Task Force}. A Just Recovery for Racial Equity in Wisconsin. tech. rep.University of Wisconsin Population Health Institute 2021.

\bibitem{paperRastEtAl2020Milwaukee}
Rast J, Martinez YC, Williams LH. {Milwaukee's Coronavirus Racial Divide: A Report on the Early Stages of COVID-19 Spread in Milwaukee County.} {\it Center for Economic Development Publications.} 2020;54.

\bibitem{paperEgedeEtAl2020WICpRace}
Egede LE, Walker RJ, Garacci E, Raymond JR. Racial/Ethnic Differences in COVID-19 Screening, Hospitalization, And Mortality In Southeast Wisconsin {\it Health Affairs. } 2020;39(11):1926-1934.

\bibitem{dataWICOVID19HistCntsumV2}
{Wisconsin Department of Health Services}. {COVID-19 Historical Data by City, Village, Town V2.} 2021.
\newblock Available at: \url{https://data.dhsgis.wi.gov/datasets/wi-dhs::covid-19-historical-data-by-city-village-town-v2}. Accessed on November 10, 2021.

\bibitem{dataCensusB03002RACE}
{U.S. Census Bureau}. {Population Estimates by Race, Table B03002, 2015-2019 American Community Survey 5-year estimates.} 2019.

\bibitem{dataCensusCartographicBoundaryCounty}
{U.S. Census Bureau}. {Cartographic Boundary Files - Shapefile: Nation-based Files - County.} 2018.

\bibitem{dataCensusCartographicBoundaryCousubdiv}
{U.S. Census Bureau}. {Cartographic Boundary Files - Shapefile: State-based Files - County Subdivisions.} 2018.

\bibitem{pageWIDHSdataInfo}
{Wisconsin Department of Health Services}. COVID-19: Frequently Asked Questions
\newblock Available at: \url{https://www.dhs.wisconsin.gov/covid-19/data-101.htm}. Accessed on July 18, 2022.

\bibitem{paperTornqvist1985relativeChange}
Tornqvist L, Vartia P, Vartia YO. How Should Relative Changes Be measured? {\it The American Statistician. } 1985;39(1):43-46.

\end{thebibliography}
\end{document}